\documentclass[11pt,a4paper]{article}
\usepackage{titlesec}
\usepackage{color}
\usepackage{fontenc}
\usepackage{listings} %inclusion of source code
\definecolor{mygreen}{rgb}{0,0.6,0}
\definecolor{mygray}{rgb}{0.5,0.5,0.5}
\definecolor{mymauve}{rgb}{0.58,0,0.82}
\definecolor{myback}{rgb}{0.92,0.95,0.95}
\usepackage{siunitx}
\usepackage{float}
\usepackage{tikz}
\usepackage[version=3]{mhchem}
\usepackage{booktabs,array}
\newcolumntype{+}{>{\global\let\currentrowstyle\relax}} \newcolumntype{^}{>{\currentrowstyle}}

\usepackage{nicefrac}
\usepackage[english]{babel} 
\usepackage[top=2.8cm,bottom=2.8cm,left=3cm,right=3cm]{geometry}  % ændre margen 
\usepackage{amsmath,amsfonts,amssymb}  % matematiske tegn og symboler
\usepackage{graphicx,subfigure}  % billeder kan lægges ind
\graphicspath{{./Figures/}{./PSTricks/}}  % billeder kan lægges i undermappen "billeder"
\usepackage{multirow}
\usepackage{appendix}
\usepackage{wrapfig}
\usepackage{braket}

\usepackage{caption3} % load caption package kernel first
\DeclareCaptionOption{parskip}[]{} % disable "parskip" caption option
\usepackage[font=small,labelfont=bf]{caption}
\usepackage[final]{pdfpages}

\usepackage{tikz}
\usepackage{float}
\usepackage{sidecap}
\sidecaptionvpos{figure}{t}
\DeclareMathOperator\erf{erf}
\begin{document}
%\begin{titlepage}
%\includepdf{Forside}
%\end{titlepage}
%
%\pagestyle{empty}
%\mbox{}
%\newpage

\pagenumbering{arabic}

\begin{center}
\textbf{{\Large Shape-preserving and unidirectional frequency conversion using four-wave mixing Bragg scattering \medskip}}
\end{center}
\begin{center}
{\large Jesper B.~Christensen$^1$, Jacob G.~Koefoed$^1$, Bryn A. Bell$^2$, \\ Colin~J.~McKinstrie$^{1,3,*}$, and Karsten Rottwitt$^1$ }
\end{center}
\begin{center}
$^1$\textit{Department of Photonics Engineering, Technical University of Denmark, 2800 Kgs. Lyngby, Denmark} \\
$^2$\textit{Clarendon Laboratory, University of Oxford, Parks Road, Oxford OX1 3PU, UK}\\
$^3$\textit{Futurewei Technologies, 400 Crossing Boulevard, Bridgewater, NJ 08807, USA} \\
$^*$\textit{colin.mckinstrie@huawei.com}
\end{center}

\noindent{\textbf{Abstract}} ---
In this work, we investigate the properties of four-wave mixing Bragg scattering in a configuration that employs orthogonally polarized pumps in a birefringent waveguide. This configuration enables a large signal conversion bandwidth, and allows strongly unidirectional frequency conversion as undesired Bragg-scattering processes are suppressed by waveguide birefringence. Moreover, we show that this form of four-wave mixing Bragg scattering preserves the (arbitrary) signal pulse shape, even when driven by pulsed pumps.

\section{Introduction}
All-optical nonlinear signal processing has opened new doors within information processing and optical communication \cite{willner2014}. By utilizing the third-order nonlinearity in optical fibers or integrated waveguides, signal processing tasks such as amplification \cite{hansryd2002,tong2011}, regeneration \cite{radic2003, guan2018}, nonlinearity mitigation \cite{hu2014,Sackey:15}, and data-format conversion \cite{guan2017}, have all been demonstrated. The common workhorse enabling these operations is four-wave mixing (FWM), which comes in different flavors depending on the required capability. 

\begin{figure}[t!]
\begin{centering}
\includegraphics[width=1\textwidth]{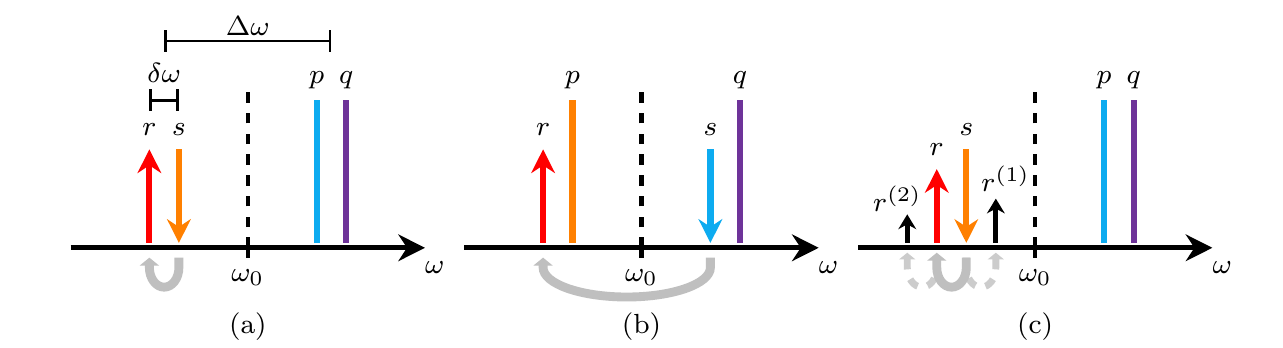}
\caption{(a) Near- and (b) distant frequency conversion by BS for fields placed symetrically around the zero-dispersion frequency $\omega_0$. The input signal $s$ is down-shifted by $\delta\omega$ in frequency to $r$ by the two pumps, $p$ and $q$. The separation between the average pump frequency and the average frequency of the input and converted signal is denoted $\Delta\omega$. The direction of the arrows indicate direction of energy flow, which may be reversed to achieve up-conversion. (c) In the near configuration, spurious Bragg scattering processes, generating the additional fields $r^{(1)}$ and $r^{(2)}$, may limit the conversion efficiency from $s$ to $r$.   \label{fig1}}
\end{centering}
\end{figure}

%regeneration, format change???, phase-sensitive parametric amplification, optical phase conjugation, time lenses, multicasting, and signal generation (Hao). 
One particular method of all-optical nonlinear signal processing is FWM Bragg scattering (BS), in which an input signal ($s$) is up- or downshifted to an output signal ($r$) by the frequency difference between two pump lasers ($p$ and $q$), as illustrated in Fig.~\ref{fig1}(a) and (b). In contrast to FWM processes such as parametric amplification and phase conjugation, which are inherently noisy, BS enables full conversion of an input signal without adding noise \cite{McKinstrie:05}. For this reason, BS has attracted attention in quantum photonics as it allows signal processing of single- or few-photon level signals \cite{mcguinness2010,Clark:13,li2016,joshi2018}. This could be useful in quantum communications for shifting quantum signals between telecom wavelengths, where the transmission loss in optical fibers is lowest, and visible wavelengths, where superior detectors, quantum memories, or single-photon sources are likely to operate \cite{li2016,tanzilli2005photonic,agha2013,Kuo:13,fernandez2013}. Shifting single photons between wavelength channels could be used for routing signals across quantum networks, or for quantum information processing with frequency encoding schemes \cite{brecht2015,ra2017,Reddy:18}. Similarly, BS has applications in classical communications and fast all-optical signal processing \cite{Myslivets:09,Zhao:16,Li:16,zhao2017}.

%In principle, BS can achieve unit conversion efficiency (CE) and introduces no noise into the signal \cite{McKinstrie:05}. {\color{red} In contrast, FWM processes such as parametric amplification or phase conjugation, the signal is amplified and a copy is created at another frequency, which quantum mechanics says must inevitably introduce excess noise. BS has attracted attention in quantum photonics because, since it is noiseless, it can be applied to single- or few- photon level signals \cite{mcguinness2010,Clark:13,li2016}.}

%

One practical challenge in BS is that of frequency \textit{unidirectionality}. As BS allows both up- and downconversion, only one of these processes should be phase matched at a time. If this is not the case, part of the signal power is up-shifted and part is down-shifted \cite{agha2013,zhao2017,agha2012,bell2016,Li:17}. This bidirectionality ($r^{(1)}$) is illustrated in Fig.~\ref{fig1}(c), which also shows the potential of cascaded conversion in the same direction ($r^{(2)}$). In order to achieve a high conversion efficiency (CE), these undesired processes must be suppressed by designing them to have large phase mismatches.

%For practical frequency conversion, the BS process should be well phase matched over the bandwidth of the signal(s), and other nonlinear processes such as parametric amplification should not be phase matched, so as to avoid introducing noise \cite{zhao2017}. In particular, since BS can in general be used for both up- and down-shifting, only one BS process should be phase matched at a time. If this is not the case, part of the signal power is up-shifted and part down-shifted \cite{agha2012,agha2013,zhao2017,bell2016,Li:17}. This bi-directionality is illustrated in Fig.~\ref{fig1}(c), which also shows the potential of the signal being converted multiple times in the same direction \cite{erkintalo2012}. Thus, in order to achieve a high conversion efficiency (CE), these undesired processes must be suppressed by designing them to have large phase mismatches.   

%This can either be realized by careful engineering of the optical density of states {\color{red} mikkel}, or by introducing mode-induced asymmetry into the phase-matching as we propose here. 

%
%see Fig.~\ref{fig1}(c), limiting the obtainable conversion efficiency (CE) in the desired direction \cite{zhao2017,bell2016,Li:17}. 
%
%This is especially a concern when the frequency shift is relatively small. Similarly, it is possible for the signal to be shifted multiple times in the same direction, and these secondary processes should be suppressed by phase-matching, or by engineering an appropriate density of states [cite Mikkel], in order to achieve full CE.

Another, yet unresolved, challenge in the framework of BS is that of achieving shape-independent and shape-preserving frequency conversion when the process is driven by short pump pulses. Intuitively, the use of pulsed pumps rather than continuous-wave (CW) pumps is advantageous in the sense that far lower average pump powers are required to achieve efficient conversion. However, in the setting of pulsed pumps, the nonlinear interaction strength varies in time, making the attainable CE strongly dependent on the temporal mode of the input signal \cite{brecht2015,mejling2012,Mckinstrie12}.

%Theoretical work has explored BS when the pumps are supplied as short pulses. This has a practical benefit in that a far lower average pump power is required for efficient conversion, but since the pump power is varying in time, it is generally not possible to achieve full conversion for an arbitrary temporal-spectral signal shape \cite{Mckinstrie12,mejling2012}.

In this work, we seek solutions to these challenges by investigating a scheme for BS where the two pump pulses are polarized on orthogonal axes of a nonlinear birefringent waveguide, or as recently demonstrated, propagate in different spatial modes of a higher-order mode fiber \cite{friis2016,Parmigiani:17}. We show that this configuration allows highly unidirectional frequency conversion with the conversion direction being controlled by the polarization of the input signal. Furthermore, the configuration enables conversion of high-bandwidth signals, and preserves the signal temporal shape even when driven by short pump pulses. Additionally, in contrast to the standard BS configuration, where the fields must be centered symmetrically around a waveguide zero-dispersion frequency (ZDF), this new scheme allows phase matching to be achieved in both the normal- or the anomalous dispersion regime.

%\cite{Christensen16,Koefoed:17}

%{\color{red} Here, we investigate a scheme for BS where the two pump pulses are polarized on orthogonal axes of a nonlinear birefringent waveguide, and are group-velocity matched to one another. The input and output signal also have orthogonal polarizations and are group-velocity matched; however the group walk-off between the pumps and the signal is large enough that the pumps completely walk over the signal within the length of the nonlinear fiber. We show that this configuration can achieve high conversion efficiency for large bandwidth signals and preserves the temporal shape of the signal, even when the pump pulses are short. Up- or down-conversion can be selected by choosing the input polarization of the signal, and phase-matching can be achieved in the normal or anomalous dispersion regime, in contrast to the standard configuration for co-polarized BS, where the frequencies used must be centred around a zero dispersion frequency (ZDF).} 

%\begin{itemize}
%\item The configuration has already been demonstrated using higher-order modes. We should probably mention this?
%\end{itemize}

\section{Standard configuration}
\subsection{Phase matching and bandwidth}
First, we discuss the standard configuration for BS, where all the fields are co-polarized, and are centered around a ZDF \cite{inoue1994}. We label the input signal $s$, the output $r$, and the pumps $p$ and $q$, as shown in Fig.~\ref{fig1}. The phase mismatch for the down-shifting case [Fig.~\ref{fig1}(a) and (b)], is given by
\begin{equation}
\Delta\beta = \beta(\omega_s) -\beta(\omega_r)+\beta(\omega_p)-\beta(\omega_q) +\gamma(P_q-P_p),
\label{mismatch}
\end{equation}
where $\beta(\omega)$ is the wavenumber at angular frequency $\omega$, $\gamma$ is the nonlinear coefficient proportional to the intensity-dependent refractive index $n_2$, and $P_{p}$ and $P_q$ are the pump powers. Notably, the nonlinear contribution to the wavenumber-matching condition cancels if the two pump powers are equal. The wavenumber as a function of frequency can be expanded as
\begin{equation}
\beta(\omega) = \beta_0 + \beta_1 \omega + \beta_3\omega^3/6 + \mathcal{O}(\omega^4),
\label{eq:phasematchstand}
\end{equation}
where $\omega$ is measured relative to the ZDF, in which case the second-order dispersion term vanishes, i.e.~$\beta_2=0$. For balanced pump powers, wavenumber matching, i.e.~$\Delta\beta=0$, is obtained by placing the fields symmetrically around the ZDW, such that $\omega_s=-\omega_p$ and $\omega_r=-\omega_q$, leading to cancellation of the odd terms in Eq.~\eqref{mismatch}. Such placement of the fields furthermore leads to group-velocity matching of $s$ to $p$, and of $r$ to $q$, as can be seen from the group slowness
\begin{equation}
\beta'(\omega)=\mathrm{d}\beta/\mathrm{d}\omega= \beta_1 + \beta_3\omega^2/2+\mathcal{O}(\omega^3).
\end{equation}
Notably, the third-order dispersion coefficient $\beta_3$ plays an important role in setting the allowed signal bandwidth, and in determining the degree to which other nonlinear processes are suppressed, or allowed. To estimate the phase-matching bandwidth, we allow $\omega_s$ to deviate from $-\omega_p$, while fixing $\delta \omega=\omega_s-\omega_r$.  Thereby, the wavenumber mismatch is given by
\begin{equation}
\Delta\beta=\frac{\beta_3\delta\omega\Delta\omega}{2}(\omega_p+\omega_s),
\end{equation}
with $\Delta\omega$ being the separation between the average frequency of the pumps and that of the input/output signal, as shown in Fig~\ref{fig1}(a). In a waveguide of length $L$, efficient conversion occurs for $\vert \Delta \beta L \vert \ll 1$, resulting in the following condition for the signal bandwidth $\Omega_s$
\begin{equation}
\Omega_s \ll \left| \frac{4}{\beta_3 \delta\omega\Delta\omega L} \right|.
\label{bandwidth}
\end{equation}

\subsection{Unidirectionality of the standard configuration}
Consider now the influence of spurious BS processes, which are sketched in Fig.~\ref{fig1}(c). These processes must be well suppressed in order to obtain a high CE in the desired direction. To quantify this, we consider the configuration for desired down-shifting from $s$ to $r$. The wavenumber mismatch for the spurious up-shifting process, $s$ to $r^{(1)}$, is in this case given by
\begin{equation}
\Delta \beta _{\mathrm{spur}} = \beta(\omega_s)-\beta(\omega_s+\delta\omega)-\beta(\omega_p)+\beta(\omega_q)=\beta_3 \omega_p \delta\omega^2,
\end{equation}
with $\omega_s+\delta\omega$ being the frequency of up-shifted light. From this, we deduce the condition for suppressing the unwanted up-shifting process ($\vert \Delta \beta_{\mathrm{spur}}  L \vert  \gg 1$)
\begin{equation}
|\beta_3 \omega_p \delta\omega^2 L|\gg 1,
\label{eq:spur1}
\end{equation}
which shows that the spurious BS process may be particularly difficult to suppress for small frequency shifts $\delta \omega$. A very similar condition applies for suppressing the process where converted light at $\omega_r$ is down-shifted a second time, i.e.~$r$ to $r^{(2)}$:
\begin{equation}
\left|\beta_3 \omega_q\delta\omega^2 L \right|\gg 1.
\label{eq:spur2}
\end{equation}
Notably, by combining the conditions for unidirectionality [Eqs.~\eqref{eq:spur1} and \eqref{eq:spur2}] with the attainable BS bandwidth [Eq.~\eqref{bandwidth}], one finds that unidirectional operation is only possible for a signal bandwidth that is much smaller than the frequency shift, i.e.~$\delta\omega \gg \Omega_s$. This could be detrimental to applications of frequency conversion for dense wavelength division multiplexing systems, where the channel separation is comparable to the bandwidth of each channel, and where spurious BS could lead to unintended interference with other channels.

%Note that if phase matching is simultaneously to allow a desired frequency conversion over a certain signal bandwidth [Eq.~\eqref{bandwidth}], and suppress spurious BS [Eqs.~\eqref{eq:spur1} and \eqref{eq:spur2}], then it is required that $\delta\omega \gg \Omega_s$.

\begin{figure}[h!]
\begin{center}
\includegraphics[scale=1]{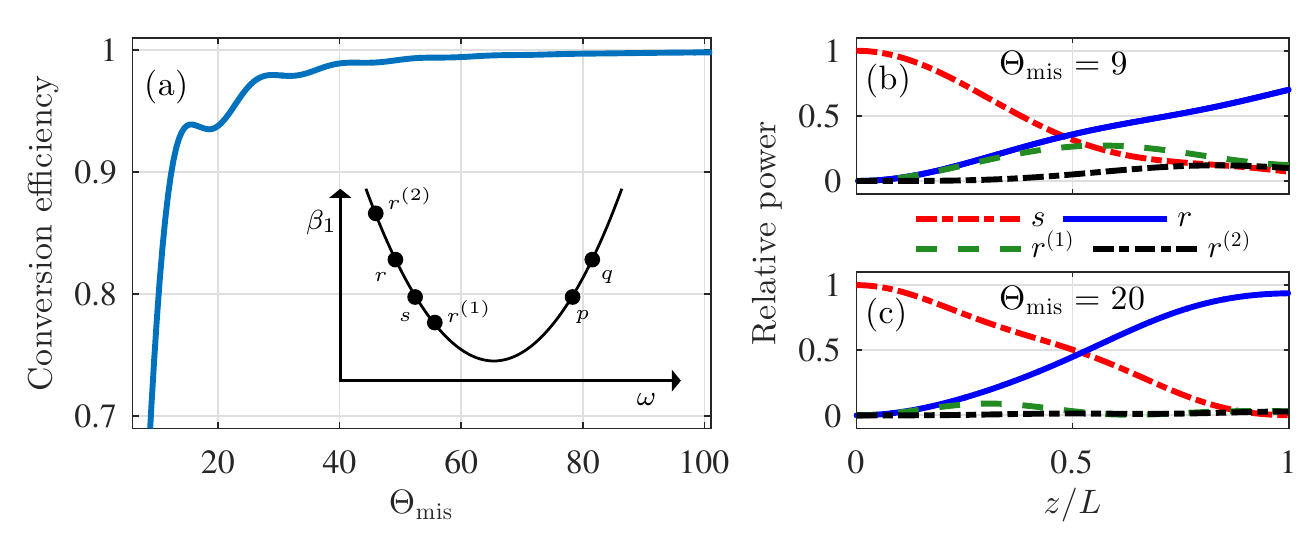}
\caption{ (a) Maximal conversion efficiency as a function of the dimensionless mismatch parameter $\Theta_\mathrm{mis} = \beta_3 \Delta \omega \delta \omega^2 L$. For small mismatches, undesired Bragg scattering modes, $r^{(n)}$, become significant, and thereby limit the conversion efficiency from $s$ to $r$. Inset shows the phase-matching diagram interpreted as a parabola in ($\omega, \beta_1$)-space and the placements of the various fields. (b) and (c) show the relative power transfer versus waveguide distance between the input signal $s$ (dashed-dotted, red) the desired output $r$ (full, blue), the undesired bidirectional output $r^{(1)}$ (dashed, green), and the cascaded converted output $r^{(2)}$ (dashed-dotted, black), for $\Theta_\mathrm{mis} = 9$ and $\Theta_\mathrm{mis} = 20$, respectively. }
\label{fig:unidirectionality}
\end{center}
\end{figure}

The effect of the wavenumber mismatches given in Eqs.~\eqref{eq:spur1} and \eqref{eq:spur2}, is quantified by solving the coupled-mode equations for the BS process, including multiple signal modes (for details, see Appendix A). Figure \ref{fig:unidirectionality}(a) shows the highest attainable CE as a function of the dimensionless product $\Theta_\mathrm{mis} = \beta_3 \Delta \omega \delta \omega ^2 L$, for desired down-conversion [Fig.~\ref{fig1}(c)]. The pump power (which is CW and balanced, $P_p = P_q$) is in all cases chosen such that a CE of unity is obtained without the inclusion of the undesired Bragg-scattering modes, $r^{(n)}$ (i.e.~$2\gamma P_p L = \pi/2$). For values $\Theta_\mathrm{mis} > 50$, the attainable CE is near unity as undesired Bragg scattering processes are strongly suppressed by the large wavenumber mismatches given in Eqs.~\eqref{eq:spur1} and \eqref{eq:spur2}.  However, for small values of the dimensionless product, the CE is limited to far below unity as a result of significant coupling to undesired signal modes. Furthermore, the CE-curve features small, and decaying, oscillations that arise due to the dynamical interaction between the multiple signal modes, as shown in Figs.~\ref{fig:unidirectionality}(b) and (c). As a consequence of the quadratic dependence of $\Theta_\mathrm{mis}$ on $\delta \omega$, small frequency shifts are particular difficult to achieve in a unidirectional fashion. As an example, for a 100-m long optical fiber with a third-order-dispersion coefficient of $\beta_3 = 1~\mathrm{ps^3}/\mathrm{km}$, a frequency shift of $\delta \omega = 2~\mathrm{THz}$ with the pumps placed $\Delta \omega = 20~\mathrm{THz}$ from the signals, yields $\Theta_\mathrm{mis} = 8$. That is, according to Fig. \ref{fig:unidirectionality}, the frequency conversion process is far from unidirectional.  This bidirectionality is an even larger hurdle in integrated waveguides, for which the length is limited to the order of centimeters, something which has been observed by multiple research groups in recent years \cite{agha2012,bell2016,Li:17}. Recently, however, it was demonstrated that the bidirectionality could be alleviated by exploiting birefringence \cite{Bell:17}, and in the following, we improve upon this concept by taking a closer look at the cross-polarized BS configuration.

\section{Cross-polarized configuration}
\subsection{Phase matching and bandwidth}
\begin{figure}[hb]
\begin{centering}
\includegraphics[scale=1]{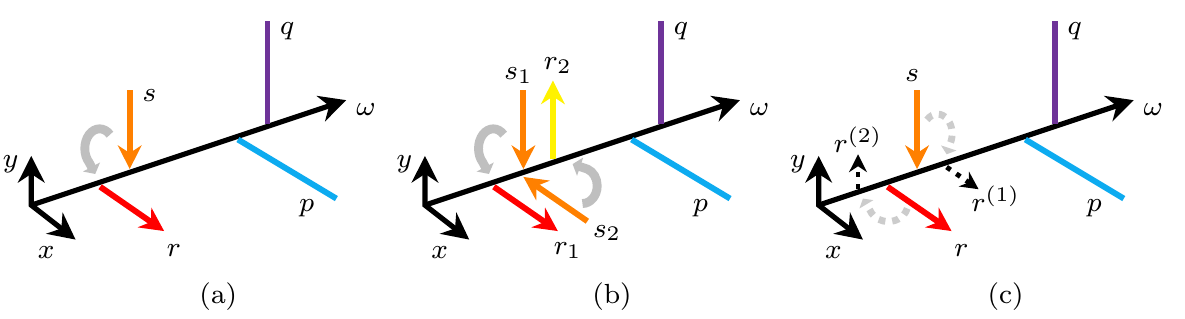}
\caption{(a) In cross-polarized BS the converted output signal is orthogonal in polarization to the input signal. (b) The direction of conversion can be controlled by setting the polarization of the input signal. (c) Undesired BS processes are suppressed by waveguide birefringence, making the cross-polarized configuration unidirectional. \label{fig3}}
\end{centering}
\end{figure}
To address the problems encountered with the standard configuration, we now consider an alternative setup where the two pumps are polarized on orthogonal axes of a birefringent nonlinear waveguide. The nonlinear waveguide is assumed to consist of a material for which the third-order susceptibility tensor takes the form $\chi^{(3)}_{ijkl} = a \delta_{ijkl} + b ( \delta_{ij} \delta_{kl} + \delta_{ik} \delta_{jl} +  \delta_{il} \delta_{jk} )$ \cite{lin2007}, where the indices $i,j,k,l$ refer to the polarization state of an electric field and $\delta$ is the Kroenecker delta function. This form of the third-order susceptibility tensor encapsulates the properties of, for example, a silica fiber or a silicon waveguide, and entails that the converted output signal is polarized orthogonally to the input signal, as shown in Fig.~\ref{fig3}(a). 

%The relevant tensor elements of the third-order susceptibility are permutations of $\chi^{(3)} _{xyxy}$, which dictates that the output signal be polarized on the orthogonal axis to the input signal, as shown in Fig.~\ref{fig3}(a).

%     The $\chi^{(3)}$ tensor then dictates that the output signal be polarized on the orthogonal axis to the input signal, as shown in Fig.~\ref{fig3}(a) for the case of down-conversion. 

%The relevant element of the $\chi^{(3)}$ tensor is three times smaller than that for the co-polarized case, which can be compensated for by increasing the power of each pump laser by a factor of three.

We consider the case where the involved fields are placed far from the waveguide ZDF, resulting in a non-zero group-velocity dispersion $\beta_2$, which may be either negative or positive. The group-velocity dispersion is assumed to dominate over higher-order dispersion terms, and is assumed identical for the principle axes of the birefringent waveguide. Expanding the wavenumber around the average frequency $\omega_\mathrm{av}$, the wavenumbers for the two principle axes are given as
%We still take the various frequencies relative to $\omega_0$, but now assume that there is a non-zero group-velocity dispersion $\beta_2$ term, which dominates over higher-order dispersion, and is identical for the two polarizations. The propagation constants for the two fibre axes are now given by:
\begin{equation}
\beta_\pm(\omega)=\beta_{0\pm}+\beta_{1\pm}\omega+\frac{\beta_2}{2}\omega^2 + \mathcal{O} (\omega^3),
\label{eq:birefringencebeta0}
\end{equation}
where $\omega$ is relative to $\omega_\mathrm{av}$, and $\pm$ indicates the slow ($+$) and fast ($-$) axis of the waveguide, respectively. If we let pump $p$ be polarized on the slow axis (and, consequently, let pump $q$ be polarized on the fast axis), we find from Eq.~\eqref{eq:birefringencebeta0} that the cross-polarized BS process is wavenumber matched, i.e.~$\Delta \beta = 0$, whenever $ \omega_q - \omega_p = \Delta \beta_1 / \beta_2$, where $\Delta \beta_1 \equiv \beta_{1+} - \beta_{1-}>0$. Notably, the process occurs, exactly as does vector-modulation instability in birefringent waveguides, for both normal ($\beta_2 >0)$ and anomalous ($\beta_2 <0$) dispersion \cite{drummond1990,amans2005}. By our convention one must choose $\omega_q > \omega_p$ for normal dispersion, and $\omega_q < \omega_p$ in the case of anomalous dispersion. 

Remarkably, in complete contrast to the standard configuration, which allows perfect phase matching only for one specific signal frequency, the cross-polarized process is wavenumber matched \textit{independently} of the signal frequency, provided the frequency separation between the pumps is chosen judiciously to $\Delta  \beta_1 / \beta_2$. Moreover, the direction of the frequency shift is determined by the polarization of the input signal as sketched in Fig.~\ref{fig3}(b). More specifically, if we let the input signal be polarized along the fast axis, we find $\omega_r = \omega_s - \Delta \beta_1 /\beta_2$ resulting in a down-shift (an up-shift) for normal (anomalous) dispersion, while if the input signal is polarized along the slow axis, we obtain  $\omega_r = \omega_s + \Delta  \beta_1 /\beta_2$ giving rise to an  up-shift (a down-shift) for normal (anomalous) dispersion. 

%Something else than this: (which is not true) Since these two processes are phase-matched simultaneously, the BS process effectively realizes the effect of a polarization beam splitter using frequency channels rather than spatial channels.  

%with the resulting group-slowness
%\begin{equation}
%\beta'_\pm(\omega)=\pm\beta_1+\beta_2\omega + \mathcal{O}(\omega^2),
%\label{eq:birefgroupslowness}
%\end{equation}

The phase-matching condition of the cross-polarized BS process furthermore leads to pairwise group-velocity matching of the pumps ($p$ to $q$), and the signals ($s$ to $r$), as can be seen by insertion of the shift $\delta \omega = \Delta \beta_1/\beta_2$ into 
\begin{equation}
\beta'_\pm(\omega)=\beta_{1\pm} + \beta_2\omega + \mathcal{O}(\omega^2).
\label{eq:birefgroupslowness}
\end{equation}
A similar group-velocity-matching condition has recently been shown to enable the generation of spectrally uncorrelated photon pairs \cite{Christensen16,mckinstrie2017,Koefoed:17}, and we consider its implications in more details in Sec.~\ref{sec:4}.

%The effect of this group-velocity matching is studied in more detail in Sec .~\ref{sec:4}.

\subsection{Unidirectionality of the cross-polarized configuration}
As for the standard configuration, we now consider the effect of spurious BS. The wavenumber mismatch of this process [$s$ to $r^{(1)}$ in Fig.~\ref{fig3}(c)], can be written as
\begin{equation}
\Delta \beta_\mathrm{spur} = 2  \Delta \beta_0 -  \Delta  \beta_1 \left( \Delta \omega + 2 \delta \omega \right) ,
\label{eq:spuriousmismatch}
\end{equation}
with $\Delta \beta_0 \equiv \beta_{0+} - \beta_{0-}$. Thus, to suppress the spurious BS, we require
\begin{equation}
\left| 4 \pi\frac{  L}{L_\mathrm{B}} -  \Delta  \beta_1 \left( \Delta \omega + 2 \delta \omega \right) L \right| \gg 1 , 
\label{eq:spuriousrequirement1}
\end{equation}
where we have introduced the mode beat length of the birefringent waveguide, $L_\mathrm{B} = 2 \pi / \Delta \beta_0$. Similarly, the suppression of the secondary BS process [$r$ to $r^{(2)} $ in Fig.~\ref{fig3}(c)], requires
\begin{equation}
\left| 4 \pi\frac{  L}{L_\mathrm{B}} -  \Delta  \beta_1 \left( \Delta \omega - 2 \delta \omega \right) L \right| \gg 1 , 
\label{eq:spuriousrequirement2}
\end{equation}
differing from Eq.~\eqref{eq:spuriousrequirement1} only by the sign in front of the frequency shift $\delta\omega$. Noteworthy, in the cross-polarized configuration, the wavenumber mismatches of the undesired BS process contain one term inversely proportional to the beat length $L_\mathrm{B}$, and another term proportional to the difference in inverse group velocity $\Delta \beta_1$. For standard birefringent fibers with beat lengths on the order of 1--10 mm \cite{noda1986}, the first term is typically orders of magnitudes larger than the second term for which $\Delta \beta_1  \approx 1~\mathrm{ps}/\mathrm{m}$ (and $\Delta \omega \approx 5$--$20~\mathrm{THz}$) \cite{Parmigiani:17,amans2005}. This is also the case in integrated birefringent waveguides, where the beat length can readily be made smaller than hundreds of micrometer at optical wavelengths \cite{kiyat2005,morichetti2007}. Hence, the condition for suppressing spurious BS effectively becomes identical to the condition for preventing polarization-mode coupling, which is exactly the primary ability of these kinds of birefringent waveguides.

\section{Shape-preserving frequency conversion}
\label{sec:4}
The use of pulsed, rather than CW, pumps, entails the need for far smaller average pump-power levels. However, this also significantly complicates the spatial-temporal dynamics, typically resulting in optimal signal CE for only a single (or few) temporal shape(s), and an altered temporal shape of the converted signal \cite{mejling2012}. In the following, we show that the cross-polarized BS configuration allows preservation of the signal temporal shape and, moreover, enables shape-independent frequency conversion. 
%{\color{red} necessary sentence?: This shape-selective nature of BS has been proposed as a means for enabling add-drop functionality of field-orthogonal single-photon temporal modes, vital in a quantum information architecture based of photonic temporal modes \cite{brecht2015,Reddy:14,christensen2015}.}

\subsection{Pump dynamics}
Let us start by considering the pump dynamics. In our configuration, phase matching dictates co-propagating pumps ($\beta_{1p} = \beta_{1q} = \beta'$) and co-propagating signals ($\beta_{1s} = \beta_{1r}$). The undepleted coupled pump equations take the form
% For simplicity, we therefore adopt a reference frame travelling along with the signals, and define $\beta' = \beta_{1p} - \beta_{1s}$, yielding the undepleted coupled pump equations 
\begin{subequations}
\begin{align}
\left(\partial_z + \beta' \partial_t   \right) & A_p =  i \gamma \left( \vert A_p \vert ^2    + \frac{2}{3} \vert A_q \vert ^2 \right) A_p ,  \\
\left(\partial_z + \beta' \partial_t   \right) & A_q =  i \gamma \left( \vert A_q \vert ^2    + \frac{2}{3} \vert A_p \vert ^2 \right) A_q  ,
\end{align}
\label{eq:pumps}%
\end{subequations}
in which the amplitudes $A_{p,q}$ are slowly varying envelopes in units of $\mathrm{W}^{1/2}$, and are assumed unaffected by intra-pulse dispersion. In Eqs.~\eqref{eq:pumps}, the terms describing cross-phase modulation contain factors of $2/3$, representing the case of orthogonally polarized fields in an isotropic material such as fused silica. This factor may be different if one considers the BS process using higher-order spatial modes \cite{friis2016,Parmigiani:17}, or in a material with an anisotropic Kerr nonlinearity such as crystalline silicon \cite{zhang2007}.  We stress, however, that what follows does not depend on the value of this prefactor. 

Optimal conversion in the third-order Kerr nonlinearity occurs when the two pump pulses are temporally matched, and we therefore consider the case where the initial pump pulses obey $A_{p_0}(0,t ) = A_{q_0}(0,t)$. With this initial condition, the pump powers are balanced, which optimizes the nonlinear interaction per total amount of pump power, and the solution to the pump evolution becomes
\begin{equation}
A_p(z,t) = A_{p_0}(t-\beta' z) \exp \left( \frac{5i\gamma}{3}  \vert A_{p_0} (t-\beta'z) \vert ^2 z  \right),
\label{eq:pumpsol}
\end{equation}
with $A_q(z,t) = A_p(z,t)$. The exponential in Eq.~\eqref{eq:pumpsol} accounts for both self- and cross-phase modulation, which contribute in the same manner as the pump pulses are group-velocity matched.

%\begin{subequations}
%\begin{align}
%\partial_z  & a_s =  i \gamma \left( 2\vert A_q \vert ^2    + \frac{2}{3} \vert A_p \vert ^2 \right) a_s +i \frac{2}{3} \gamma A_p^* A_q a_i \\
%\partial_z  & a_i =  i \gamma \left(2 \vert A_p \vert ^2    + \frac{2}{3} \vert A_q \vert ^2 \right) a_i +i \frac{ 2}{3} \gamma A_p A_q^* a_s 
%\end{align}
%\end{subequations}
%Write signal and idler equations on matrix form? 
%

\subsection{Signal dynamics}
Consider now the Heisenberg-picture coupled-mode equations for the signal-mode operators. In the signal reference frame, so that now $\beta' = \beta_{1p} - \beta_{1s}$, we have 
\begin{equation}
\partial_z \begin{bmatrix}
a_s(z,t) \\
a_r(z,t)
\end{bmatrix} = \mathbf{M}(z,t) \begin{bmatrix}
a_s(z,t) \\
a_r(z,t)
\end{bmatrix},
\label{eq:signaleqs}
\end{equation}
where the system matrix $\mathbf{M}$ is given by
\begin{equation}
\mathbf{M}(z,t)= i \gamma \begin{bmatrix}
2\vert A_q(z,t) \vert ^2    + \frac{2}{3} \vert A_p(z,t) \vert ^2  &  \frac{2}{3} A_p^*(z,t) A_q(z,t) \\
 \frac{2}{3} A_p(z,t) A_q^*(z,t) & 2 \vert A_p(z,t) \vert ^2    + \frac{2}{3} \vert A_q(z,t) \vert ^2
\end{bmatrix}.
\end{equation}
In our case, we may define $A(z,t)  \equiv A_p(z,t) = A_q(z,t)$, recasting $\mathbf{M}$ into the simple form
\begin{equation}
\mathbf{M}(z,t) = \frac{2 i \gamma}{3} \begin{bmatrix}
4 & 1 \\
 1 & 4
\end{bmatrix}\vert A(z,t) \vert ^2 ,
\end{equation}
in which the matrix part is no longer spatially dependent. As a result, $\mathbf{M}$ commutes with itself at different spatial positions, i.e.~$[\mathbf{M}(z',t),\mathbf{M}(z'',t)] = 0$, and therefore, Eq.~\eqref{eq:signaleqs} is solved by \cite{blanes2009}
\begin{equation}
\begin{bmatrix}
a_s(z,t) \\
a_r(z,t)
\end{bmatrix} = \begin{bmatrix}
G_{ss} (z,z_0,t) & G_{sr} (z,z_0,t)   \\
G_{rs} (z,z_0,t) & G_{rr} (z,z_0,t)
\end{bmatrix} \begin{bmatrix}
a_s(z_0,t) \\
a_r(z_0,t)
\end{bmatrix} =  \mathbf{G}(z,z_0,t) \begin{bmatrix}
a_s(z_0,t) \\
a_r(z_0,t)
\end{bmatrix},
\end{equation}
where the $2\times 2$ matrix transfer function is of the form $\mathbf{G}(z,z_0,t) =\exp \left[ \int_{z_0} ^z \mathrm{d}z' \mathbf{M}(z',t) \right]$.  This fact allows us to directly write down the solution as
\begin{align}
\mathbf{G}(L,0,t) =  \exp \left[ 4 i  \xi(t)  \right] 
\times \begin{bmatrix}
\cos \left[\xi(t)  \right] &  i\sin \left[  \xi(t) \right] \\
i\sin \left[ \xi(t) \right] & \cos \left[  \xi(t)  \right]
\end{bmatrix},
\label{eq:Gtransformation}
\end{align}
where we have defined the effective interaction strength
\begin{equation}
\xi(t ) = \frac{2 \gamma}{3}  \int _{z_0=0} ^{z=L} \mathrm{d}z' \vert A_0 (t-\beta' z') \vert ^2 .
\label{eq:cumpump}
\end{equation}
The exponential in Eq.~\eqref{eq:Gtransformation} encompasses the combined effects of cross-phase modulation from the two pump pulses, whereas the matrix describes the time-dependent beam-splitter-like transformation, which is typical for nonlinear frequency-conversion processes \cite{Reddy:14}. As is apparent from Eq.~\eqref{eq:cumpump}, the CE efficiency of a time slice $t_k$ depends on the interaction strength experienced by that time slice according to $\sin ^2[ \xi(t_k) ]$. Therefore, in general, the converted signal is a distorted version of the input signal. However, if the input signal experiences a complete temporal collision with the pumps (a complete walk-off), then the CE is time-independent (CW-like), and the signal shape is preserved. Remarkably, this shape-preserving property, which is unique to our configuration, holds for both arbitrary input signal- and pump shapes.

%The CE of a time slice $t_k$ is directly related to the cumulative overlap of the pumps with that time slice according to $\sin ^2[ 2 \gamma  \xi(L,0,t_k) / 3]$. Moreover, as a direct consequence of the group-velocity-matching condition, converted photons are perfectly correlated in time with corresponding input photons. } {\color{red} This stands in contrast to usual configurations, for which $s$ and $r$ are group-velocity \textit{mis}matched, potentially leading to converted photons being temporally (and spectrally) uncorrelated with the signal input photons \cite{eckstein2011,mejling2012}. 
%\begin{itemize}
%\item $t_k \rightarrow t_k$ upon conversion. Strikingly different from other configurations where each input time slice is connected to a continuum of output time slices (due to signal-idler walk-off).  
%\end{itemize}

\subsection{Examples with Gaussian pumps}
We now consider a few examples, and, for simplicity, consider Gaussian-shaped input pump pulses of the form 
\begin{equation}
A_0 (t) = \left( \frac{E} { \pi^{1/2} \tau} \right)^{1/2} \exp\left[ - (t+t_0)^2/(2 \tau^2 )\right],
\label{eq:gaussianpump}
\end{equation}
where $E$ is the pulse energy and $\tau$ is the pulse duration (related to the root-mean square width $T_\mathrm{RMS}$ according to $\tau = \sqrt{2} T_\mathrm{RMS}$). The parameter $t_0$ determines the initial pulse center in our reference frame, and henceforth $t_0 = \beta' L/2$ is used to ensure that the pump pulses are centered on $t=0$ after a propagation distance of $z=L/2$. With the pulse shape in Eq.~\eqref{eq:gaussianpump}, one can readily show that Eq.~\eqref{eq:cumpump} takes the explicit form 
\begin{equation}
\xi (t) = \frac{\gamma E}{3 \beta' } \left[ \erf \left( \frac{t}{\tau } + \frac{\zeta}{2} \right) - \erf \left( \frac{t}{\tau } - \frac{\zeta}{2}\right) \right],
\label{eq:CEgaussian}
\end{equation}

where $\erf$ is the error function, and we have defined the dimensionless walk-off parameter $\zeta = \beta' L/\tau$, which quantifies the degree of walk-off between the pumps and a time slice of the signal. Notably, if a time slice $t_k$ experiences a full collision with the pumps, then $\xi(t_k) =2 \gamma  E/(3 \beta')$, and the CE (of this time slice) is then only dependent on the interaction strength $\gamma E/ \beta'$, which is a product of the nonlinearity $\gamma$, the pump peak power $E/\tau$, and the walk-through distance $\tau/\beta'$.

\begin{figure}[h]
\begin{center}
\includegraphics[scale=1]{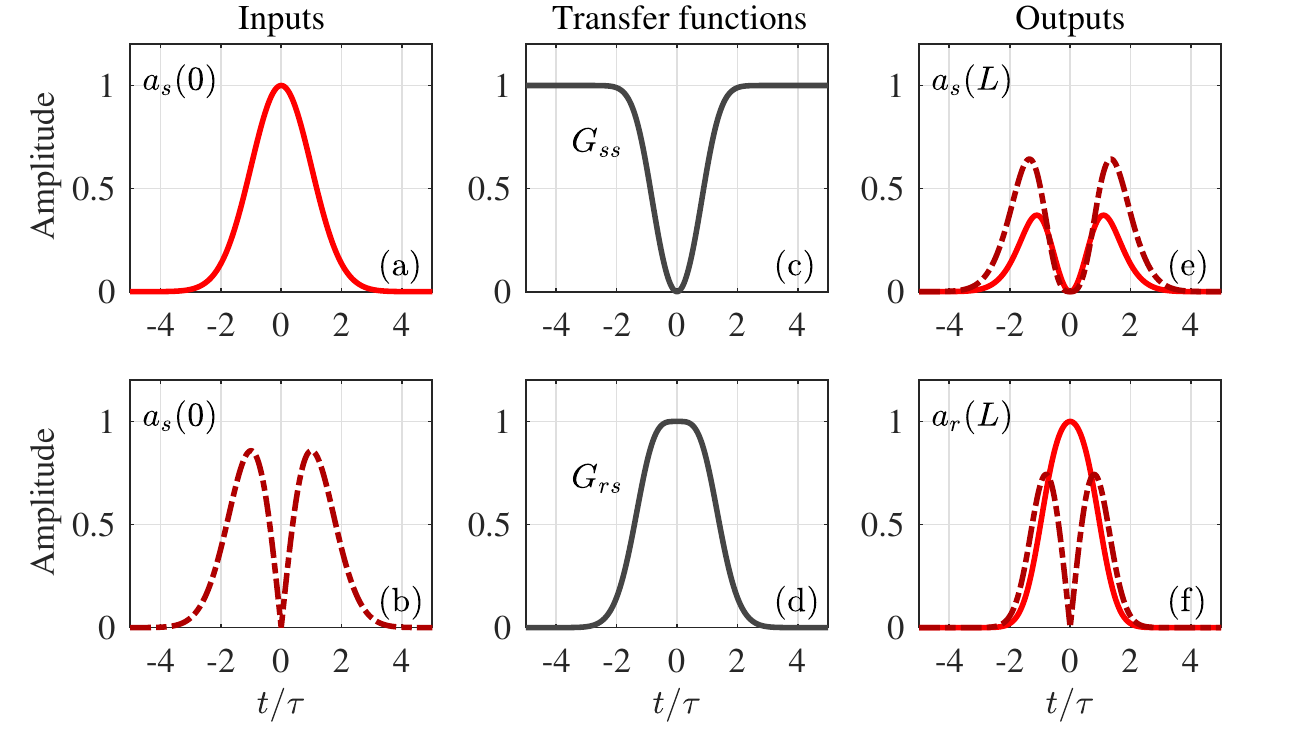}
\caption{ (a) Gaussian- and (b) first-order Hermite-Gaussian signal $s$ inputs, which multiplied by (c) the self-transfer function $G_{ss}$ and (d) the cross-transfer function $G_{rs}$, yields, (e) the remaining signal $s$ outputs, and (f) the converted signal $r$ outputs, respectively. The walk-off parameter, $\zeta = 2$, does not enable a full collision between the pumps and the signal resulting in temporally localized conversion. }
\label{fig4}
\end{center}
\end{figure}

Figure \ref{fig4} illustrates the conversion dynamics in the case of $\zeta = 2$ for two different signal input pulse shapes; (a) a Gaussian input, and (b) a first-order Hermite-Gaussian input.  The interaction strength is chosen such that the center of the signal, i.e. $t=0$, is fully converted as illustrated with the transfer functions in Figs.~\ref{fig4}(c) and (d). However, a value of $\zeta = 2$ only allows for a moderate degree of walk-off, resulting in a temporally localized conversion as shown in Figs.~\ref{fig4}(e) and (f) for the remaining $s$-output and converted $r$-output, respectively. Notably, as the Hermite-Gaussian input signal contains only a small part of its energy around the pulse center, the total CE is only $53\%$ compared to $85\%$ for the Gaussian input signal. Moreover, the converted signal is temporally narrower than the input signal, but is spectrally broadened due to the chirp received as a result of cross-phase modulation from the pumps. 

\begin{figure}[h]
\begin{center}
\includegraphics[scale=1]{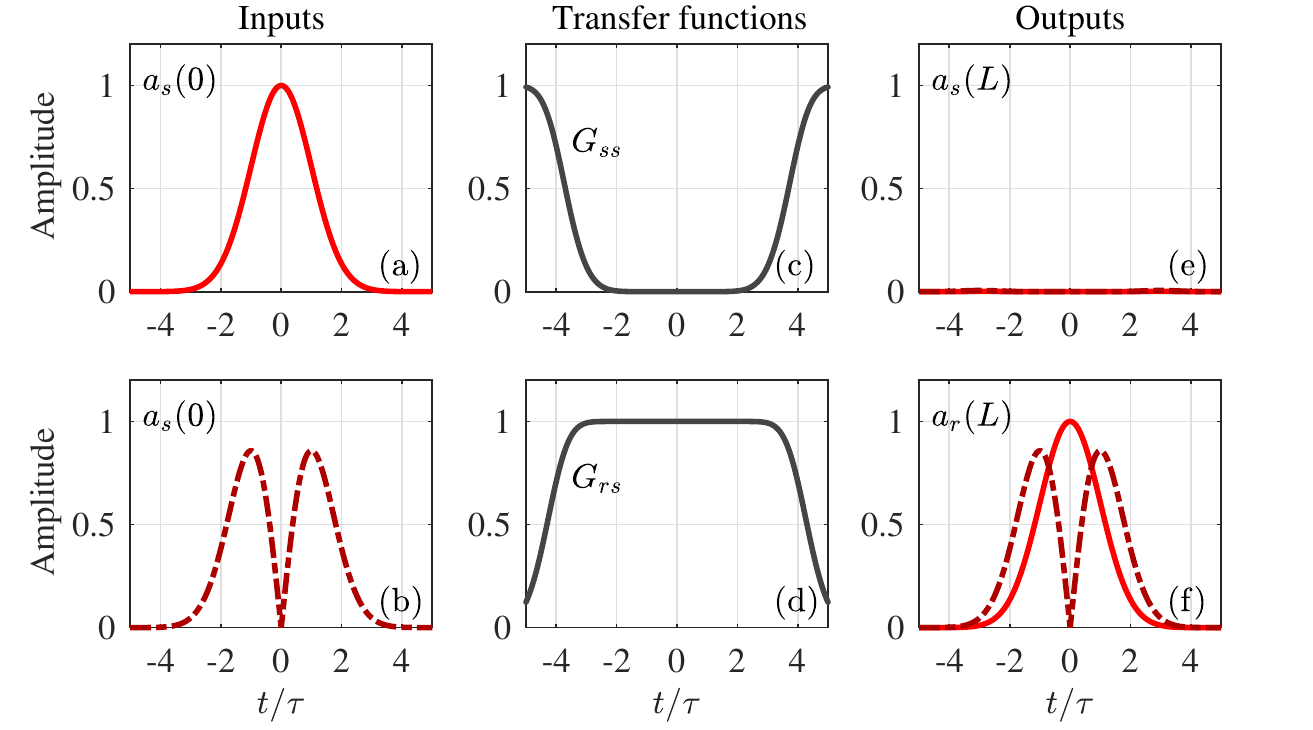}
\caption{(a) Gaussian- and (b) first-order Hermite-Gaussian signal $s$ inputs, which multiplied by (c) the self-transfer function $G_{ss}$ and (d) the cross-transfer function $G_{rs}$, yields, (e) the remaining signal $s$ outputs, and (f) the converted signal $r$ outputs, respectively. The walk-off parameter, $\zeta = 8$, in this case enables a full collision between the pumps and the signal, resulting in shape-preserving conversion of both considered input signals.   }
\label{fig5}
\end{center}
\end{figure}

%{\color{red} Figure \ref{fig4} shows how input signals in the forms of (a) a Gaussian pulse and (b) a first-order Hermite-Gaussian pulse are (e) transmitted and (f) converted for $\zeta = 2$, i.e~a small degree of pump-signal walk-off.   The interaction strength ($\gamma E/ \beta'$) is chosen such that the time slice $t=0$ is fully converted as seen from the transfer functions (c) $G_{ss}$ and (d) $G_{rs}$. However, in this case, the interaction is temporally localized resulting in optimal conversion for only the central part of the input signal. As the Hermite-Gaussian contains only a small part of its energy around the pulse center, the CE is only xxx compared to yyy for the Gaussian signal input. (Insert interpretation for quantum signals?) Additionally, as a result of the localized temporal interaction, the converted signal is temporally narrower than the input signal, but spectrally broadened due to the chirp received from cross-phase modulation.  }

% pump-power independent phase (the phase is constant)

The case of $\zeta = 8$ is shown in Fig.~\ref{fig5}, using the same input signal pulses as in Fig.~\ref{fig4}. Now the walk-off allows a full pump-signal collision, giving rise to transfer functions that are approximately constant within the signal duration, as seen in Figs.~\ref{fig5}(c) and (d). This results in CEs of practically unity for both the Gaussian and the first-order Hermite Gaussian input signals, demonstrating that the configuration enables conversion of arbitrary temporal shapes with high efficiency. Moreover, irrespective of the signal input shape, this temporal shape is preserved in the frequency conversion process, as seen from Fig.~\ref{fig5}(f).

Finally, it is highly instructive to make a comparison between the proposed pulsed scheme and the corresponding CW pumped configuration. In the CW pumped regime, the CE is simply determined by the product $\gamma P_{\mathrm{cw}} L$. On the other hand, when employing pulsed pumps, the conversion process typically becomes complicated, resulting in the CE being strongly dependent on the exact temporal shape, and timing, of the input signal. However, in this configuration, when the pumps are allowed to completely `scan' through the signal, the CE [see Eqs.~\eqref{eq:Gtransformation} and \eqref{eq:CEgaussian}] becomes time-independent. Thus, one can think of this configuration as being quasi CW-like, with a conversion efficiency being determined solely by the interaction strength $\gamma E/\beta'$. Hence, in comparison to the CW case, the power-length product $P_{\mathrm{cw}} L$ is replaced by $E/\beta'$. Thereby, to maintain a given CE moving from the CW regime to the pulsed regime, the peak power should merely satisfy $P_p = E/(\pi^{1/2} \tau) = P_{\mathrm{cw}} \beta' L /(\pi^{1/2} \tau) \approx 5 P_{\mathrm{cw}} $, where we have used that $\zeta \approx 8$ for a full collision with comparable signal and pump durations (see Fig.~\ref{fig5}).

%CW-like..

%It is highly instructive

%{\color{red} Consider now Fig.~\ref{fig5}, which, like Fig.~\ref{fig4}, illustrates the BS process, but for $\zeta = 8$. In this case, the pumps traverse completely through the signal giving rise to transfer functions that are constant in both amplitude and phase, as seen in Figs.~\ref{fig5}(c) and (d).} Choosing again the appropriate interaction strength for full conversion at the center of the signal, we find CEs of {\color{red} yyy2 and xxx2} for the Gaussian and first-order Hermite Gaussian, respectively. Thus, in contrast to the earlier example, here the two different signal inputs are converted with practically the same efficiency. Moreover, irrespective of the signal input shape, this temporal shape is preserved in the BS conversion process.  

\section{Discussion}
In recent years, there have been proposals for using BS to perform all-optical switching and logic operations \cite{Zhao:16,Li:16}. The cross-polarized configuration adds an extra degree of freedom; polarization; and allows parallel operation of both polarization modes, which in the process are converted in opposite directions. Moreover, this scheme bears potential for enabling very high signal conversion bandwidths as described in Sec.~3.1, where we assumed a simple waveguide-dispersion profile with polarization-independent group-velocity dispersion. Although this can be the design target, in practice, the waveguide dispersion takes a form which is only approximately described by Eq.~\eqref{eq:birefringencebeta0}, and one may need to include a polarization-dependent group-velocity dispersion ($\beta_{2\pm}$) and/or higher-order-dispersion coefficients to rigorously model the phase-matching condition. This alteration naturally limits the signal conversion bandwidth, which, in general, decreases with increasing departure from Eq.~\eqref{eq:birefringencebeta0}.

Another consequence of the considered dispersion, is that the cross-polarized BS process is phase matched only when the two orthogonally polarized pump waves are placed in frequency such that they are (approximately) group-velocity matched. This entails that the size of the frequency shift is determined by the group birefringence and the group-velocity dispersion (recall, $\delta \omega = \Delta \beta_1/\beta_2$). While this suggests that any given waveguide can only enable a certain designed frequency shift, the inclusion of a polarization-dependent group-velocity dispersion [in which case, $\delta \omega = \Delta \beta_1/\beta_{2,av}$, with $\beta_{2,av} = (\beta_{2+} + \beta_{2-})/2$] or third-order dispersion (still, $\delta \omega = \Delta \beta_1/\beta_2$, but with $\beta_2$ now being frequency dependent) opens the possibility for fine-tuning the frequency shift by tuning the pump frequencies (which changes the average frequency $\omega_\mathrm{av}$ around which the wavenumbers are expanded).

%This suggests that any given waveguide can only enable a certain designed frequency shift, 

%Thus, unless one makes use of methods to tune the birefringence \cite{kerbage2002}, any given waveguide can only enable a certain designed frequency shift.  
Finally, it is worth noticing that the cross-polarized BS process is designed to occur when the fields are placed far away from a waveguide ZDF. For this reason, the considered process is, in comparison to the standard BS configuration, to a lesser degree accompanied by other parasitic nonlinear processes such as parametric amplification. This is a highly desired feature, as it enables less noisy operation.

%3/4-1 page of discussion, on the following:
%\begin{itemize}
%\item Only specific frequency shift. (check)
%\item Placement far from ZDW, allows cleaner operation. (check)
%\item Nonlinear polarization switch. (check)
%\item CW-versus-pulsed comparison.
%\item Application to quantum.
%\item Dispersion assumption. (check)
%\end{itemize}

%\begin{itemize}
%\item Would it be sufficient to try and model previously used fibers? We would need to find data in other papers [Amans: no effective index difference (but we could calculate approximately from $\Delta \beta_1$), Drummond and Rotherberg contain beat length, GV mismatch, and GVD]
%\end{itemize}

\section{Conclusion}
We have investigated the properties of a four-wave mixing Bragg-scattering configuration, which employs cross polarized pumps in a birefringent nonlinear waveguide.  Phase matching of this process, which can be achieved in both the anomalous or the normal dispersion regimes, occurs when the two pumps are placed in frequency such that they are group-velocity matched. The cross-polarized configuration has four distinct advantages compared to the standard co-polarized Bragg-scattering configuration; (i) it allows a large signal bandwidth, which is not limited by the size of the frequency shift; (ii) the direction of conversion (up or down) is controlled by the polarization of the input signal; (iii) conversion is entirely unidirectional as undesired Bragg-scattering processes are suppressed by waveguide birefringence; and (iv) the pairwise group-velocity matching (pump-to-pump and signal-to-signal) enables shape-preserving frequency conversion of an arbitrary signal input temporal waveform.

%\begin{itemize}
%\item nonlinear polarization switch with potentially broad bandwidth
%\item unidirectional Bragg scattering
%\item due to group-velocity matching of the fields, the converted signal has the same temporal form as the input signal, and the conversion process is independent of the input signal if the pumps fully scan the input signal.
%\end{itemize}

\section*{Acknowledgments}
DFF Sapere Aude Adv. Grant NANO-SPECs and DFF (Grant No. 4184-00433).

\appendix

\section{Coupled-mode equations with multi-level Bragg scattering}
To model the standard co-polarized BS process including multiple signal modes we introduce multiple additional converted signal frequencies according to $\omega_{r^{(n)}} = \omega_s +  \delta \omega  (n+1)/2$ for $n$ odd (up-converted fields), and $\omega_{r^{(n)}} = \omega_s - \delta \omega (n/2+1) $ for $n$ even (down-converted fields), where $\delta \omega = \omega_q - \omega_p > 0$, [see also inset of Fig.~\ref{fig:unidirectionality}(a)]. For each of these frequencies, the corresponding wavenumber is obtained from  Eq.~\eqref{eq:phasematchstand}, resulting in the wavenumber mismatches 
\begin{equation}
\Delta \beta_{j\rightarrow k} = \beta (\omega_j) - \beta (\omega_k)    \pm \left( \beta(\omega_p) - \beta (\omega_q) \right),
\end{equation}
where $+$ is chosen for $\omega_j > \omega_k$ and $-$ is chosen for $\omega_j < \omega_k$. Note, that this construction entails $\Delta \beta_{j\rightarrow k} = - \Delta \beta_{k\rightarrow j}$, as required. With the input fields placed such that $\omega_s = - \omega_p$, with frequencies measured relative to the waveguide ZDF, we have, for example, $\Delta \beta_{s\rightarrow r} = 0$, but $\Delta \beta_{s\rightarrow r^{(1)}} \neq 0$. The BS process only allows coupling between signal fields separated by $\delta \omega$, and thus the set of first-order coupled ordinary differential equations takes the form 
\begin{subequations}
\begin{align}
&\partial_z  a_s =  2i \gamma \left( \vert  A_p \vert ^2    +  \vert A_q \vert ^2 \right) a_s + 2 i \gamma \left( A_p^* A_q a_r +A_p A_q^* a_{r^{(1)}} \mathrm{e}^{i \Delta \beta_{r^{(1)} \rightarrow s}}  \right)  ,  \\
&\partial_z  a_r =  2i \gamma \left( \vert  A_p \vert ^2    + \vert A_q \vert ^2 \right) a_r + 2 i \gamma \left( A_p^* A_q a_{r^{(2)}}\mathrm{e}^{i \Delta \beta_{r^{(2)} \rightarrow r}}  +A_p A_q^* a_s  \right)  ,  \\
&\partial_z  a_{r^{(1)}} =  2i \gamma \left( \vert  A_p \vert ^2    + \vert A_q \vert ^2 \right) a_{r^{(1)}}+ 2 i \gamma \left( A_p^* A_q a_s\mathrm{e}^{i \Delta \beta_{s \rightarrow r^{(1)}}}  +A_p A_q^* a_{r^{(3)}} \mathrm{e}^{i \Delta \beta_{r^{(3)} \rightarrow r^{(1)}}}   \right)  ,  \\
&\partial_z  a_{r^{(2)}} =  2i \gamma \left( \vert  A_p \vert ^2    + \vert A_q \vert ^2 \right) a_{r^{(2)}}+ 2 i \gamma \left( A_p^* A_q a_{r^{(4)}}\mathrm{e}^{i \Delta \beta_{r^{(4)} \rightarrow r^{(2)}}}  +A_p A_q^* a_r \mathrm{e}^{i \Delta \beta_{r \rightarrow r^{(2)}}}   \right)  , \\
& \hphantom{\partial_z  a_r^{(2)} a} \vdots   \nonumber
\end{align}
\label{eq:pumpsap}%
\end{subequations}
where the vertical dots present the equations for $a_{r^{(n>2)}}$. In the numerical simulations used to create Fig.~\ref{fig:unidirectionality}, we included, for each value of $\Theta_\mathrm{mis}$, $N$ additional fields so that negligible power transfer was observed to the most detuned fields, i.e.~$n = N-1$ and $n = N$. 

\bibliographystyle{unsrt}
\bibliography{biblo}
%
%\begin{thebibliography}{99}
%\bibitem{Gisin07}N.~Gisin and R.~Thew, ``Quantum communication,'' Nat.~Photonics \textbf{1}, 165--171 (2007).

%\end{thebibliography}

\end{document}